\documentclass[twocolumn,showpacs,amsmath,amssymb,superscriptaddress]{revtex4}
\usepackage{graphicx}
\usepackage{dcolumn}
\usepackage{bm}
\preprint{APS/123-QED}

\begin{document}

\title{Photon-number selective group delay in cavity induced transparency}
\author{Gor Nikoghosyan$^{1,2}$ and Michael Fleischhauer}
\affiliation{Department of physics and research center OPTIMAS, University of
Kaiserslautern, Erwin-Schr\"{o}dinger-Strasse, D-67663 Kaiserslautern,
Germany\\
$^{\textit{2}}$ Institute of Physical Research, 378410, Ashtarak-2, Armenia}

\begin{abstract}
We show that the group velocity of a probe pulse in an ensemble of
$\Lambda$-type atoms driven by a quantized cavity mode depends on the quantum state of the input probe pulse.
In the strong-coupling regime of the atom-cavity system the probe group delay is photon number selective.
This can be used to spatially separate the single photon from higher photon-number components of a few-photon
probe pulse and thus to create a deterministic single-photon source.
\end{abstract}

\maketitle


One of the major practical challenges in implementing photon-based quantum cryptography
and network quantum computing is the controlled, deterministic generation of single
photon pulses. In the present paper we propose a scheme where the group delay of a probe
pulse in a medium with electromagnetically induced transparency (EIT) is made quantum state
depended. This can be used to build a Fock-state quantum filter and thus to create single-photon pulses
on demand.

EIT is an interference effect
where the optical properties of a probe field are modified by the presence
of a strong, and thus usually classical, coupling field
\cite{Harris-Physics-Today-1997, Fleischhauer-RMP-2005}. Due to destructive interference
induced by the coupling field an otherwise opaque
resonant medium becomes transparent in a narrow spectral region. The transparency is accompanied by a
substantial reduction of the group velocity which can be controlled by the
intensity of the coupling field \cite{Hau-Nature-1999}.
We here consider the case when
the classical driving field is replaced by a quantized cavity mode. If the corresponding vacuum Rabi-frequency
is sufficiently large, i.e. if the cavity-atom system is in the strong coupling regime, already an
empty cavity will induce transparency for the probe field \cite{Field-PRA-1993}.
Furthermore a weak probe pulse will induce cavity enhanced Raman scattering into the resonator mode. Under appropriate
conditions almost all excitations will be transferred to the cavity mode and thus its photon number
distribution will be a copy of that of the input probe field. As the probe-field group
velocity in the EIT medium depends on the photon number of the drive field, different photon number components
of the probe pulse will experience different group delays. We will show that the differential group delay between the single- and higher photon number components can be made large enough to fully separate them spatially during propagation.

It is well-known that stong coupling of a cavity mode to atomic dipoles can give rize to nonlinearities
that are sufficiently large to induce interactions on the few photon level and thus can be employed
for photonic quantum gates and determinsitic single-photon sources \cite{cavity-QED-Kimble,Parkins-PRL-1993,Kuhn-PRL-2000,Kuhn-PRL-2007}.
However, to achieve strong coupling in the
optical domain remains a major technical challenge.
The necessity to perform input-optput operations at the same mode for which
a high quality factor is needed is a major obstacle.
Furthermore for photon transport optical frequencies are preferrable while
strong coupling is much easier achievable for $\mu$-waves.
The strong requirements of cavity QED can be
relaxed when photons interact for sufficiently long time in a nonlinear medium. E.g. it has
been suggested that photons propagating in an coherently driven, optically thick medium
under conditions of EIT can mutually induce nonlinear phase shifts due to the combination of
a strongly reduced group velocity and resonantly enhanced nonlinearities. \cite{Schmidt-Imamoglu-OL, Harris-PRL-1999, Lukin-PRL-2000, Lukin-RMP-2002}. To reach the
single-photon level in these scheme it is however necessary to let the pulses co-propagate
for large distances and to confine the light beams
transversally to a radius below the wavelength. The physical mechanism of
the present proposal is very different from both approaches. Here strong coupling is
required for the coupling field rather than
the probe field. As the latter does not need to be coupled in or out, the
probe bandwidth is not limited by the high Q value of the cavity mode. Furthermore
the frequency of the coupling field does not need to be in the optical domain.
E.g. using molecules or Rydberg atoms an optical probe transition can be
combined with a driving transition in the $\mu$-wave regime for which very large
cavity couplings have been achieved, e.g. using strip-line resonators \cite{Schoellkopf}.

\begin{figure}[hbtp]
 \includegraphics[width=9cm] {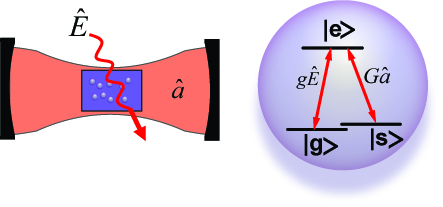}
 \caption{(color online) Schematic diagram of the system: A quantized probe field $\hat E$
interacts with an ensemble of $\Lambda$-type atoms driven by a cavity mode $\hat a$ (left)
in a Raman type coupling scheme (right).
 }
  \label{fig:system}
 \end{figure}

Let us consider a medium consisting of an ensemble of three-level atoms with a $%
\Lambda $ configuration interacting with two quantum fields (Fig.\ref%
{fig:system}). The probe $\hat{E}$ resonantly couples the
transition $\left\vert g\right\rangle $-$\left\vert e\right\rangle $ with
a coupling strength $g$ and propagates along the $z$ axis.
The $\left\vert s\right\rangle $-$\left\vert
e\right\rangle $ transition is driven by a cavity field $%
\hat{a}$.   The corresponding vacuum Rabi frequency is denoted as $G$.
The interaction time is assumed to be much shorter than the lifetime
of a photon in the cavity mode and thus we may regard the cavity as lossless.
All atoms of the medium are initially in the
ground state $\left\vert g\right\rangle $. The combination of the probe field
with the cavity vacuum field resonantly drives a Raman transition. When
the probe field enters the medium it induces Raman scattering resulting in the
creation of a collective ``spin'' excitation of the $|g\rangle -|s\rangle$ transition
accompanied by a simultaneous emission of photons into the cavity, provided the cavity coupling
is sufficiently large \cite{Field-PRA-1993, Parkins-PRL-1993, Kuhn-PRL-2000, Kuhn-PRL-2007}.

We assume, that our medium is (in the absence of EIT) optically thick for the probe field and that all
atoms are in the strong interaction regime with the cavity mode
\cite{cavity-QED-Kimble, cavity-QED-Haroche, cavity-QED-Walther}.
Although this assumption may seem on first glance rather strong, such systems are state-of-the
art technology.
Several groups have recently reported experimental results on
Bose-Einstein condensates coupled to a cavity in  the single-atom strong
coupling regime \cite{Esslinger-Nature-2007, Reichel-Nature-2007, Stamper-Kurn-2007, Vuletic-arxiv-09}.
Furthermore the cavity mode does not
have to be an optical field even if the probe field is.
E.g. one can use polar molecules instead of atoms \cite{Yelin-PRA-2008}, which can
be strongly coupled to stripline microwave resonators \cite%
{DeMille-PRL-2006}, \cite{Petrosyan-PRL-2008}, which in principle enables
very high optical depths simultaneously with a strong interaction regime.

We will now analyze the propagation of a quantum probe pulse in such a system.
To simplify the discussion we neglect the spatial structure of the cavity mode, i.e.
the coupling strength of the cavity mode is assumed to be the same for all atoms of the ensemble.
In this case  the Hamiltonian of our system  reads in the rotating wave
approximation
\begin{eqnarray}
\hat{H} &=&-\hbar N\int\limits_{0}^{L}\frac{\mathrm{d}z}{L}\Bigl[ g\hat{E} \label{hamiltonian}%
\left( z,t\right) \hat{\sigma}_{eg}\left( z,t\right) +\\
&&\qquad\qquad + \hat{a}\left( t\right)
G\hat{\sigma}_{es}\left( z,t\right) + H.c.\Bigr], \nonumber
\end{eqnarray}
where, $\hat{\sigma}_{mn}(z,t)
=\dfrac{1}{N_z}\sum\limits_{j=1}^{N_{z}}\left\vert m\right\rangle _{j}\left\langle
n\right\vert _{j}$ are the atomic flip operators averaged over a small volume around position $z$
containing $N_{z}$ atoms. $\hat{E}%
\left( z,t\right) $ is the dimensionless slowly-varying operator of the
probe field,  $\hat{a}\left(
t\right) $ denotes the annihilation operator for the cavity mode, $N$ is the total number of atoms in the
ensemble and $L$ is the length of the atomic cloud.

%
%
The propagation of the probe field is described by the Maxwell equation for
the slowly varying amplitude
\begin{equation}
\left( \frac{\partial }{\partial t}+c\frac{\partial }{\partial z}\right)
\hat{E}(z,t)=igN\hat{\sigma}_{ge}(z,t),  \label{Maxwell}
\end{equation}
while the cavity mode obeys the Heisenberg equation%
\begin{equation}
\frac{{\rm d} \hat{a}(t)}{{\rm d} t}=iN\int\limits_{0}^{L}\frac{\mathrm{d}z}{%
L}G^{\ast }\hat{\sigma}_{se}(z,t).  \label{cavity}
\end{equation}

In the strong-coupling regime the vacuum Rabi frequency $G$ of the cavity field
is large compared to the coupling strength $g$ of the probe field, i.e. $G\gg g$,
and already the cavity vacuum prepares a dark state, i.e. causes EIT \cite{Field-PRA-1993}.
If the spectrum of the input probe
pulse is within the EIT linewidth, the system will remain in the dark state and thus the
population of the excited state can be neglected. Then from eq.(\ref{hamiltonian})
one can find for the dynamics of population of the metastable states
\begin{eqnarray}
&& \frac{\partial }{\partial t}\hat{\sigma}_{gg}(z,t)= ig\hat{E}^{\dagger}(z,t)\hat{\sigma}_{ge}(z,t)-ig\hat{\sigma}_{eg}(z,t)\hat{E}(z,t), \nonumber\\
&& \frac{\partial }{\partial t}\hat{\sigma}_{ss}(z,t)= iG\hat{a}^{\dagger}(t)\hat{\sigma}_{se}(z,t) - iG\hat{\sigma}_{es}(z,t)\hat{a}(t) \label{population}
\end{eqnarray}

By substituting (\ref{Maxwell}) and (\ref{cavity}) into (\ref{population}) we find that the total
number of photons in the system, i.e. in the cavity field $\hat n(t)=\hat a^\dagger(t)\hat a(t)$
and in the probe field $\hat n_p(t)=\frac{1}{L}\int\limits_{0}^{L}\hat{E}^{\dagger}\left(
z,t\right) \hat{E}\left( z,t\right) \mathrm{d}z $ is fixed by the input and output flux
of probe photons.
\begin{eqnarray}
 \frac{{\rm d}}{{\rm d}t}\Bigl(\hat n+ \hat n_p \Bigr) = \frac{c}{L}\left(
\hat{E}^{\dagger}\left( 0\right) \hat{E}%
\left( 0\right) -\hat{E}^{\dagger}\left( L\right) \hat{E}\left( L
\right)\right)\label{photon_number}
\end{eqnarray}
The right-hand side\ of (\ref{photon_number}) is the probe field photon flux difference at the input
and output of the medium. The dynamical equations for the remaining atomic operators are
\begin{eqnarray}
&&\frac{\partial \hat{\sigma}_{ge}}{\partial t} = -\Gamma\hat{\sigma}_{ge}+iG\hat{a}\hat{\sigma}_{gs}+ig\hat{E}\left( \hat{\sigma}_{gg}-\hat{\sigma}_{ee}\right) + \hat{F}_{ge}, \quad \label{atom_new1} \\
&&\frac{\partial \hat{\sigma}_{se}}{\partial t} = -\Gamma\hat{\sigma}_{se}+ig\hat{E}\hat{\sigma}_{sg}+iG\hat{a}\left( \hat{\sigma}_{ss}-\hat{\sigma}_{ee}\right) + \hat{F}_{se},  \label{atom_new2} \\
&&\frac{\partial \left(\hat{a}\hat{\sigma}_{gs}\right)}{\partial t} = iG^{\ast }\hat{a}\hat{\sigma}_{ge}\hat{a}^{\dagger} - ig\hat{a}\hat{E}\hat{\sigma}_{es} + \hat{F}_{gs}\label{atom_new3} + \nonumber\\
&& \qquad\qquad\quad+\hat{a}\hat{\sigma}_{gs}\frac{\partial \hat{a}}{\partial t}\hat{a}^{\dagger} - \frac{\partial \hat{a}}{\partial t} \hat{a}^{\dagger}\hat{a}\hat{\sigma}_{gs} .
\end{eqnarray}
Here $\Gamma$ is the relaxation rate of the upper level and it is assumed that the decoherence
of the lower level transition is negligible on the time scale of interest. We assume that the atomic ensemble is initially prepared in the collective ground state $|g\rangle$. Then by taking into account that the population of the excited and spin states cannot exceed the number of probe photons ($n_{p}$) in the system one
can give bounds for the diagonal operators
$\hat{\sigma}_{ee}, \hat{\sigma}_{ss} \leq {n_{p}}/{N}$ and $\hat{\sigma}_{gg} \geq 1 - {n_{p}}/{N}$.
The number of the photons shall be much smaller than the number of atoms, i.e. $\epsilon\equiv \sqrt{n_{p}/N}\ll 1$.
Keeping only terms proportional to $\epsilon$ and
neglecting terms ${\cal O}(\epsilon^2)$ in eqs.  (\ref{atom_new1}), (\ref{atom_new2}) and  (\ref{atom_new3}), yields $\hat{\sigma}_{gg} \approx \hat{1}; \quad \hat{\sigma}_{ss} = \hat{\sigma}_{ee} = \hat{\sigma}_{es} = 0$ as well as
\begin{eqnarray}
&&\frac{\partial }{\partial t}\hat{\sigma}_{ge}=-\Gamma\hat{\sigma}_{ge}+iG\hat{a}\hat{\sigma}_{gs}+ig\hat{E} + \hat{F}_{ge},  \label{atom2} \\
&&\frac{\partial }{\partial t}\left(\hat{a}\hat{\sigma}_{gs} \right)= iG^{\ast }\hat{a}\hat{\sigma}_{ge}\hat{a}^{\dagger} + \hat{F}_{gs} + \nonumber\\
&&\qquad\qquad\quad +\hat{a}\hat{\sigma}_{gs}\frac{\partial \hat{a}}{\partial t}\hat{a}^{\dagger} - \frac{\partial \hat{a}}{\partial t} \hat{a}^{\dagger}\hat{a}\hat{\sigma}_{gs}. \label{atom3}
\end{eqnarray}
Note that neglecting terms ${\cal O}(\epsilon^2)$ is always justified
in the limit of a large atomic ensemble and does not mean neglecting nonlinear interactions.
The latter result here from the coupling to the cavity mode, which, as will be shown in the
following, has a photon number distribution that is an exact copy of that of the
input probe field in the limit of large atom number.
The characteristic length of the probe pulse $T$ is typically large compared to the upper level
relaxation time ($\Gamma T \gg 1  $) and thus the time derivative in eq. (\ref{atom2}) can be neglected.
If the spectrum of the probe pulse
lies within the EIT transparency window $\Delta\omega_{\rm EIT}$, i.e. if furthermore
\begin{equation}
T\, \gg\, \frac{1}{\Delta\omega_{\rm EIT}} = \frac{\Gamma}{G^2 \sqrt{\rm OD}}\qquad {\rm OD}\, \equiv\,  \frac{L}{l_{\rm abs}}
 \label{adiabaticity},
\end{equation}
where $l_{abs}={c\Gamma}/{g^{2}N}$ is the resonant absorption length of the medium in the absence of EIT, and OD the optical depth, $\hat\sigma_{ge}$ can be adiabatically eliminated, and the Langevin noise
operators can be disregarded \cite{Fleischhauer2002, Nikoghosyan2005}.
In this adiabatic limit
the atomic dynamics is governed by the equations
\begin{eqnarray}
&& \quad G\hat{a}\hat{\sigma}_{gs}+g\hat{E} = 0, \qquad\qquad \label{atom4} \\
&&\frac{\partial }{\partial t}\left(\hat{a}\hat{\sigma}_{gs} \right)= iG^{\ast }\hat{a}\hat{\sigma}_{ge}\hat{a}^{\dagger}+ \hat{a}\hat{\sigma}_{gs}\frac{\partial \hat{a}}{\partial t}\hat{a}^{\dagger} - \frac{\partial \hat{a}}{\partial t} \hat{a}^{\dagger}\hat{a}\hat{\sigma}_{gs} \qquad\label{atom5}
\end{eqnarray}
Combining eqs. (\ref{atom4}), (\ref{atom5}) and (\ref{Maxwell}) one
finally arrives at the following
propagation equation of the probe field
\begin{equation}
\frac{\partial \hat{E}}{\partial t} +
\frac{G^2}{g^{2}N}\hat{a}\left( \frac{\partial \hat{E}}{\partial t}+c\frac{\partial \hat{E}}{\partial z} \right)\hat{a}^{\dagger} - \hat{E}\frac{\partial \hat{a}}{\partial t}\hat{a}^{\dagger} + \frac{\partial \hat{a}}{\partial t} \hat{a}^{\dagger}\hat{E}=0 \label{prop1}
\end{equation}
The first terms in eq.(\ref{prop1}) describe a probe field propagation with a cavity dependenr group velocity.
The last two terms describe a dynamical reduction of the probe amplitude if the
cavity field changes in time, i.e. during the periods of entering and
leaving the medium.
Since we are not interested in these transients and in order to simplify the
discussion we will disregard these terms in the following. Taking into account that
cavity and probe operator commute we arrive at an operator-valued group velocity
\begin{equation}
\hat{v}_{\rm gr}=c\dfrac{G^{2}(\hat n +1)}{G^{2}(\hat n +1) +g^{2}N}.
\label{vgroup}
\end{equation}
$\hat{v}_{\rm gr}$ depends on the number of photons $\hat n = \hat a^\dagger \hat a$ in the cavity. On the other hand
the cavity photon number is determined by
the number of probe photons. The relation between these quantities can  be
derived from (\ref{atom4}). This yields:
\begin{equation}
G^ 2\, \hat n\left( t\right)  =g^{2} N
\int\limits_{0}^{L}\frac{\mathrm{d}z}{L}\hat{E}^{\dagger}\left( t,z\right) \hat{E}%
\left( t,z\right) , \label{photon_number2}
\end{equation}
which is nothing else than the condition for the system to remain in the
dark state.
If the number of atoms is large, such that
\begin{equation}
Ng^{2}\gg G^{2},  \label{cond1}
\end{equation}
the number of probe photons in the medium is at all times negligible as compared to the
number of photons in the cavity mode. Thus
\begin{eqnarray}
\hat n(t) -\hat n(0)\approx \frac{c}{L}\int\limits_{0}^{t}\!\mathrm{d}\tau \left[
\hat{E}^{\dagger}\left( 0\right) \hat{E}\left( 0
\right) -\hat{E}^{\dagger}\left( L\right) \hat{E}\left( L\right)\right].
\label{photon_number3}
\end{eqnarray}
In the case of an initially empty cavity the propagation of the probe in the medium
is entirely determined by the number of probe photons that have entered the medium.
This is the main idea of the present paper.

In the case of a classical driving field and thus a c-number group
velocity the spatial length of the probe pulse inside the medium is determined by the
product of the group velocity and the pulse duration $T$: $L_{\rm probe}=v_{\rm gr}T$.
In the present case the group velocity is different for the different photon-number
components of the pulse. It increases for increasing photon number. For
the sake of simplicity we assume that the spatial length of the highest relevant
photon-number component of the probe is smaller
than the medium length ($L_{\rm probe}\ll L$) and the cavity is initially empty.
In this case the whole probe pulse can be
loaded into the medium. Once the probe pulse has fully entered, the
number of photons in the cavity field becomes equal to the overall number of
probe photons at input $\hat n=\hat{n}_{p,{\rm in}}=\dfrac{c}{L}\int\limits_{0}^{t}\hat{E}^{\dagger}\left( 0,\tau
\right) \hat{E}\left( 0,\tau \right) \mathrm{d}\tau $. Now by making use of
inequality (\ref{cond1}) one finds the following solution of the propagation
equation (\ref{prop1}).
\begin{eqnarray}
\hat{E}\left( z,t\right)  &=&\hat{E}\left( 0,t-\frac{z}{\hat{v}_{\rm gr}}\right),
\label{solution} \\
\hat{v}_{gr} &=&c\frac{G^{2}(\hat{n}_{p, {\rm in}}+1)}{g^{2}N}.  \notag
\end{eqnarray}
One sees that the group velocity of the probe depends on the initial
number of photons in the pulse, thus different Fock components of the probe
pulse will propagate with different group velocities.

\begin{figure}[hbtp]
 \includegraphics[width=9cm] {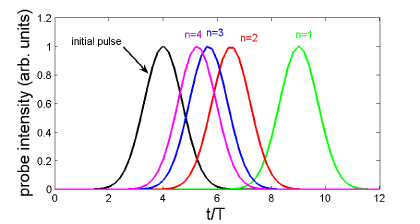}
 \caption{(color online) Spatial separation of an initial probe pulse into Fock state
components.
 }
  \label{fig:sep}
\end{figure}

To be specific let us assume that the probe field
is initially in a single-mode superposition of Fock states, i.e. the initial state of the
probe can be expressed as $\left\vert \psi \left(t\right) \right\rangle
=\sum\limits_{n=1}^{\infty }\alpha _{n}f\left( t\right) \left\vert
n\right\rangle \,$, where the common function $f\left( t\right) $ describes the shape
of the probe field before entering the medium. Note that this function is the same for all
Fock components corresponding to a single (pulsed) mode.
After propagating through the medium the state is
$\left\vert \psi(L,t) \right\rangle = \sum\limits_{n=1}^{\infty}\alpha _{n}f\left(t-\dfrac{L}{l_{abs}}\dfrac{\Gamma}{(n+1)G^2}\right) \left\vert
n\right\rangle $. Thus different components of the probe will be spatially
separated (Fig.\ref{fig:sep}). This separation is larger
for Fock components with smaller number of photons. Specifically the delay between
components with $m$ and $m+1$ photons after propagating over distance $L$ is given by
\begin{equation}
 \Delta\tau_{m} = \dfrac{L}{l_{abs}}\dfrac{1}{(m+1)(m+2)}\dfrac{\Gamma}{G^2}. \label{tau}
\end{equation}
%

Important practical limitations of the present scheme result from dissipation in the form of cavity damping and spontaneous emission. Cavity damping comes into play as soon as the cavity mode is excited, and causes a
violation of (\ref{photon_number}). It can be neglected if
\begin{equation}
 n_{p, in}\kappa T \ll 1. \label{damping}
\end{equation}
where $\kappa$ is the cavity decay rate. Spontaneous decay of the upper state can be disregarded if the interaction is
completely adiabatic (\ref{adiabaticity}).
Hence the technique works optimal if both conditions, (\ref{adiabaticity}) and (\ref{damping}), are satisfied
which can be realized only in case of strong coupling $ G^2\gg \kappa \Gamma$. However we emphasize
that since $ G\gg g$ no strong coupling is required on the probe transition.
In order to separate the single photon component
from components with larger number of excitations the delay time $\Delta\tau_1$ has to be of
order of $T$. Thus by combining (\ref{adiabaticity}) and (\ref{tau}) we find that in order to effectively separate single photon component
the following condition has to be satisfied
\begin{equation*}
 1 \lesssim \dfrac{\Delta\tau_1}{T}=\dfrac{L}{l_{abs}}\dfrac{\Gamma}{6 G^{2} T} \ll \sqrt{\dfrac{L}{l_{abs}}},
\end{equation*}
and thus a medium with large optical depth $L/l_{abs}$ is required.

Experimental realistic parameters of the cavity are $G\approx10MHz$, $\Gamma=3MHz$, $\kappa\approx1MHz$. State-of-the-art technology enables loading more than $10^5$ atoms into the cavity and thus allows to create optically thick atomic cloud with optical depth OD=$L/l_{abs} \gtrsim 10$. Under these experimental conditions proof of principle experiments should be realizable with few $ \mu sec$ pulses. Better results can in principle be obtained if microwave cavities and polar molecules are used.

In the present paper we have discussed the propagation of a weak quantum pulse
in an atomic $\Lambda$-type medium in Raman resonance with a quantized mode
of a resonator. We have shown that in this scheme the group velocity of
the probe pulse depends on the quantum state of the cavity mode, specifically
on the number of photons. In the limit of a strong cavity coupling and
for a sufficiently large optical depth of the medium for the probe field,
the cavity photon statistics is determined by the photon number
of the initial probe field. Under these conditions the action of the medium
on the probe pulse can be described in terms of a photon-number dependend
group velocity. The differential group delay between different Fock-state component
can become large enough to spatially separate the single photon from higher
photon number components of the probe. An important application of the latter
effect is a quantum state filter which can be employed to build a deterministic
single-photon source. The main advantage of the present scheme as compared to
other cavity-QED set-ups lies in the separation of the input-output mode and
the cavity mode including the possibility of very different frequencies.

G.N. acknowledges the support by the Alexander von Humboldt Foundation and by the Armenian
National Science and Education Fund.

\end{document}